\documentclass[twocolumn,showpacs,pra,floatfix]{revtex4}

\usepackage{graphicx}
\usepackage{bm}
\usepackage{psfrag}
\usepackage{amsmath}
\usepackage{amssymb}
\usepackage{latexsym}
\usepackage{exscale}

\newcommand{\ten}[1]{\mbox{\textbf{\textsf{#1}}}}

\begin{document}

\title{Local-field correction to one- and two-atom van der Waals
interactions}

\author{Agnes Sambale}

\author{Stefan Yoshi Buhmann}

\author{Dirk-Gunnar Welsch}

\affiliation{Theoretisch-Physikalisches Institut,
Friedrich-Schiller-Universit\"{a}t Jena,
Max-Wien-Platz 1, 07743 Jena, Germany}

\author{Marin-Slodoban Toma\v{s}}

\affiliation{Rudjer Bo\v{s}kovi\'{c} Institute, P.O. Box 180,
10002 Zagreb, Croatia}

\date{\today}

\begin{abstract}
Based on macroscopic quantum electrodynamics in linearly and causally
responding media, we study the local-field corrected van der Waals
potentials and forces for unpolarized ground-state atoms placed within
a magnetoelectric medium of arbitrary size and shape. We start from
general expressions for the van der Waals potentials in terms of the
(classical) Green tensor of the electromagnetic field and the
atomic polarizability and incorporate the local-field correction by
means of the real-cavity model. In this context, special emphasis is
given to the decomposition of the Green tensor into a medium part
multiplied by a global local-field correction factor and, in the
single-atom case, a part that only depends on the cavity
characteristics. The result is used to derive general formulas for the
local-field corrected van der Waals potentials and forces.
As an application, we calculate the van der Waals potential between
two ground-state atoms placed within magnetoelectric bulk material.
\end{abstract}

\maketitle


\section{Introduction}
\label{sec_intro}

Van der Waals (vdW) forces are non-vanishing forces between neutral,
unpolarized objects that arise as a consequence of quantum
ground-state fluctuations of the electromagnetic field and the
interacting objects. The first theory of vdW forces within the
framework of full quantum electrodynamics (QED) goes back to
the pioneering work of Casimir and Polder in 1948 \cite{Casimir}.
Using a normal-mode expansion of the quantized electromagnetic field,
they derived the vdW force on a ground-state atom in front of a perfectly
conducting plate as well as that between two ground-state atoms in
free space.

Whereas the normal-mode technique requires specification
of the geometry of the system at a very early point of the
calculation, alternative approaches based on linear response theory
allow for obtaining geometry-independent results. Employing the
dissipation-fluctuation theorem, geometry-independent expressions for
the vdW potentials of one and two ground-state atoms in terms of the
(classical) Green tensor of the macroscopic Maxwell equations
and the polarization of the atom can be derived
\cite{McLach1963,McLach1964,Agarwal1975,Mahanty,Mahantybook,%
wylie}---results that can be confirmed
\cite{Buhmann2004,Safari06,review2006} within the framework of
exact macroscopic QED in linear, causal media
\cite{Dung1998,Scheel01,Dung2003}.

The abovementioned methods apply to atoms in free space only, because
the local electromagnetic field acting on an atom situated in a medium
differs from the macroscopic one due to the influence of the local
medium environment of the atom. Within macroscopic QED, one approach
to overcome this problem and to account for the local-field correction
is the real-cavity model \cite{Onsager1936}, where the guest atom is
assumed to be surrounded by an small, empty, spherical cavity. The
change of the electric field due to the presence of the cavity is
then expressed via the associated Green tensor that characterizes
the local environment of the guest atom.

The real-cavity model has been applied to the spontaneous decay of an
excited atom in a bulk medium \cite{Scheel1999,Dung2003} and at the
center of a homogeneous sphere \cite{Tomas2001}. The local-field
corrected decay rate for the latter case was found to be the
uncorrected rate multiplied by a global factor and shifted by a
constant term. It was suggested that the result, reformulated in terms
of the associated Green tensor, remains valid for arbitrary
geometries. This was proven correct in Ref.~\cite{Ho2006}, where it
it is shown that the local-field corrected Green tensor can under very
general assumptions be decomposed into the sum of the uncorrected
Green tensor modified by a factor and a translationally invariant
term.

Such a decomposition of the Green tensor is not only appropriate to
obtain the abovementioned results for the rate of spontaneous decay.
It can also be used to calculate local-field corrected vdW potentials.
A first hint was given in Ref.~\cite{Tomas2006}, where the local-field
corrected vdW interaction of two ground-state atoms placed in adjacent
semi-infinite magnetoelectric media is studied.

Van der Waals interactions play an important role in various fields,
where often the interacting objects are embedded in media so that
local-field effects may be expected to play a major role. For example,
in colloid science the mutual vdW attractions of colloidal particles
suspended in a liquid influence the stability of such suspensions
\cite{Derjaguin1994, Tadros1993,Gregory1993}. In particular,
clustering of particles (flocculation) may occur unless the suspension
is  sufficiently balanced by electrostatic repulsive forces
\cite{Lawler1993, Thomas1999}. Examples of vdW interactions of
microobjects embedded in media can also be found in biology
\cite{Israelachvili1974,Nir1976}, such as cell--cell,
cell--substratum, cell--virus, and cell--vesicle interactions
\cite{Nir1976}, where the investigations are usually based on the
works of Lifshitz \cite{Lifshitz1955,ReviewLifshitz} and Hamaker
\cite{Hamaker}. So, it has been studied whether the theory of vdW
interactions can explain the phenomenon of biomolecular organization
\cite{Parsegian1973}. In particular, electrostatic and vdW forces
between a protein crystal and a molecule in solution have been
(theoretically) compared \cite{Grant94}, and the non-additivity of vdW
interactions between two layers within a multilayer assembly such as
lipid--water has been investigated \cite{Podgornik2006}. Needless to
say that vdW interactions represent only one effect among others and
that the systems are usually very complex, involving a broad range of
issues such as surface charges, simultaneous existence of many
different molecules, flexibility of biological macromolecules,
fluidity of membranes, and various types of non-covalent forces, e.g.,
vdW, electrostatic, solvation, steric, entropic, and structural ones
\cite{Oss1988, Israelachvili2005}.

In the present paper, single- and many-atom vdW interactions of
ground-state atoms embedded in media are studied within the
framework of macroscopic QED, where the local-field correction is
included via the real-cavity model. The paper is organized as
follows. In Sec.~\ref{sec_basic}, basic formulas for the vdW
potentials of ground-state atoms are reviewed. Section~\ref{sec_real}
is concerned with the real-cavity model, where a general decomposition
of the local-field corrected Green tensors entering the single- and
two-atom vdW potentials is presented. The results are combined in
Sec.~\ref{sec_local} with the general formulas for the vdW potential
as given in Sec.~\ref{sec_basic}. Closed expressions for the
local-field corrected single- and many-atom vdW potentials are
presented and briefly discussed. As an application, the vdW potential
is calculated for two atoms placed within a magnetoelectric bulk
material. Some concluding remarks are given in Sec.~\ref{sum}.


\section{Basic equations}
\label{sec_basic}

The vdW potential of a (non-magnetic) atom can be identified with the
position-dependent part of the shift of the (unperturbed) eigenenergy,
which arises from the interaction of the atom with the medium-assisted
electromagnetic field in the ground state. In leading-order
perturbation theory, the vdW potential of a ground-state atom $A$ at
position $\mathbf{r}_A$ in the presence of a linearly responding,
dispersing, and absorbing medium, can be given in the form
\begin{equation}
\label{potone}
U(\mathbf{r}_A)=\frac{\hbar \mu_0}{2\pi}
\int _0^{\infty}\mathrm{d}u\, u^2 \mathrm{Tr}
\left[\bm{\alpha}_A(iu)\!\cdot\!
\ten{G}^{(1)}(\mathbf{r}_A,\mathbf{r}_A,iu)\right]
\end{equation}
(for a derivation within the framework of macroscopic QED, see, e.g.,
Ref.~\cite{Buhmann2004}). Here,
\begin{equation}
\label{alpha}
\bm{\alpha}_A(\omega)=\lim _{\epsilon\to 0+}
\frac{2}{\hbar}\sum _k
\frac{\omega^{k}_A\mathbf{d}^{0k}_A\mathbf{d}^{k0}_A}
{(\omega^{k}_A)^2-\omega^2-i\omega\epsilon}
\end{equation}
is the atomic ground-state polarizability tensor in lowest
non-vanishing order of perturbation theory (see, e.g.,
Ref.~\cite{fain}), with $\omega^{k}_{A}$ $\!=$ $\!(E_k^A$ $\!-$
$\!E_0^A)/\hbar$ and $\mathbf{d}^{ij}_{A}$ $\!=$
$\!\langle i|\hat{\mathbf{d}}_A|j\rangle$, respectively, being the
(unperturbed) atomic transition frequencies and the atomic
electric-dipole transition matrix elements. Further, the scattering
part of the (classical) Green tensor
$\ten{G}^{(1)}(\mathbf{r},\mathbf{r'},\omega)$ of the electromagnetic
field accounts for scattering at inhomogeneities of the medium. Recall
that the Green tensor of the entire system can be decomposed according
to
\begin{equation}
\label{bulkscatter}
\ten{G}(\mathbf{r},\mathbf{r'},\omega)=
\ten{G}^{(0)}(\mathbf{r},\mathbf{r'},\omega)
+\ten{G}^{(1)}(\mathbf{r},\mathbf{r'},\omega)
\end{equation}
[$\ten{G}^{(0)}(\mathbf{r},\mathbf{r'},\omega)$, bulk part]. In
particular for locally responding, isotropic, magnetoelectric
matter, the Green tensor satisfies the differential equation
\begin{equation}
\label{DGLgreen}
\left[\bm{\nabla}\times \frac{1}{\mu(\mathbf{r},\omega)}\bm{\nabla}
\times\,-\frac{\omega^2}{c^2}\varepsilon(\mathbf{r},\omega)\right]
\ten{G}(\mathbf{r},\mathbf{r'},\omega)=
\delta(\mathbf{r}-\mathbf{r'})\ten{I}
\end{equation}
($\ten{I}$, unit tensor) together with the boundary condition
\begin{equation}
\ten{G}(\mathbf{r},\mathbf{r'},\omega)\rightarrow 0
\quad
\mbox{for }\left|\mathbf{r}-\mathbf{r'}\right|
\rightarrow \infty.
\end{equation}
The (conservative) force $\mathbf{F}(\mathbf{r}_A)$
associated with the potential (\ref{potone})
can be obtained according to
\begin{equation}
\label{forceone}
\mathbf{F}(\mathbf{r}_A)= -\bm{\nabla}_{\!\!A}U(\mathbf{r}_A).
\end{equation}

Next, consider two neutral, unpolarized (non-mag\-netic) ground-state
atoms $A$ and $B$ in the presence of a linearly responding medium. The
force acting on atom $A$ due to the presence of atom $B$ can be
derived from the two-atom potential obtained in fourth-order
perturbation theory as (cf.~Ref.~\cite{Safari06})
\begin{multline}
\label{pottwo}
U(\mathbf{r}_A,\mathbf{r}_B)=-\frac{\hbar\mu_0^2}{2\pi}
\int _0^{\infty} \mathrm{d}u \,u^4\\
\times \mathrm{Tr}\left[\bm{\alpha}_A(iu)\!\cdot\!
\ten{G}(\mathbf{r}_A, \mathbf{r}_B,iu)\!\cdot\!
\bm{\alpha}_B(iu)
\!\cdot\!\ten{G}(\mathbf{r}_B,\mathbf{r}_A,iu)\right].
\end{multline}
Note that in contrast to Eq.~(\ref{potone}), the full Green
tensor $\ten{G}(\mathbf{r}_A,\mathbf{r}_B,\omega)$ comes into play.
The force $\mathbf{F}_{A(B)}(\mathbf{r}_A,\mathbf{r}_B)$ on the atom
$A(B)$ due to the presence of the atom $B(A)$ can be calculated
according to
\begin{equation}
\label{forcetwo}
\mathbf{F}_{A(B)}(\mathbf{r}_A,\mathbf{r}_B)
=-\bm{\nabla}_{\!\!{A(B)}}U(\mathbf{r}_A,\mathbf{r}_B).
\end{equation}
Equation~(\ref{pottwo}) can be generalized to the $N$-atom potential
$U(\mathbf{r}_1,\dots, \mathbf{r}_{N})$. In particular, for
spherically symmetric atoms ($\bm{\alpha}_{A}$ $\!=$
$\!\alpha_{A}\ten{I}$), one finds \cite{buhmann_born}
\begin{multline}
\label{many}
U(\mathbf{r}_1,\dots, \mathbf{r}_{N})\\=
\frac{(-1)^{N-1}\hbar\mu_0^{N}}{(1+\delta_{2N})\pi}
\int _0^\infty \mathrm{d}u\,u^{2N}
\alpha_1(iu)\cdots\alpha_N(iu)\\
\times {\cal{S}}\mathrm{Tr}
\left[\ten{G}(\mathbf{r}_1,
\mathbf{r}_2,iu)\cdots
\ten{G}(\mathbf{r}_{N},\mathbf{r}_1,iu)\right],
\end{multline}
where the symbol $\cal{S}$ introduces symmetrization with respect to
the atomic positions $\mathbf{r}_1,\dots,\mathbf{r}_{N}$, and the
corresponding force on the atom $J$ ($J$ $\!=$ $\!1,2,\ldots,N$)
is
\begin{equation}
\mathbf{F}_{J}(\mathbf{r}_1,\ldots,\mathbf{r}_J,\ldots,\mathbf{r}_{N})
=-\bm{\nabla}_{\!\!J}U(\mathbf{r}_1,
\ldots,\mathbf{r}_J,\ldots,\mathbf{r}_{N}).
\end{equation}


\section{Real-cavity model}
\label{sec_real}

Let us consider a guest atom in a host medium. In this case, the
local electromagnetic field at the position of the atom, i.e., the
field the atom interacts with is different from the macroscopic
field. This difference can be taken into account by appropriately
correcting the  macroscopic field to obtain the local field relevant
for the atom--field interaction. Within macroscopic electrodynamics,
a way to introduce local-field corrections is offered by the
real-cavity model. Under the assumption that the guest atom (at
position $\mathbf{r}_A$) is well separated from the neighboring host
atoms, in the real-cavity model it is assumed that the guest atom
is located at the center of a small free-space region, which in the
case of an isotropic host medium is modeled by a spherical cavity of
radius $R_\mathrm{c}$. Accordingly, the system of the unperturbed host
medium of permittivity $\varepsilon(\mathbf{r},\omega)$ and
permeability $\mu(\mathbf{r},\omega)$ is replaced with the system
whose permittivity $\varepsilon_\mathrm{loc}(\mathbf{r},\omega)$
and permeability $\mu_\mathrm{loc}(\mathbf{r},\omega)$ read
\begin{equation}
\label{loc}
\varepsilon_\mathrm{loc}(\mathbf{r},\omega),
\mu_\mathrm{loc}(\mathbf{r},\omega)
=\begin{cases}
1&\hspace*{-1.5ex}\mbox{if }
\left|\mathbf{r}-\mathbf{r}_A\right|< R_\mathrm{c},\\
\varepsilon(\mathbf{r},\omega),
\mu(\mathbf{r},\omega)
&\hspace*{-1.5ex}\mbox{if }\left|\mathbf{r}-
\mathbf{r}_A\right|\ge R_\mathrm{c}.
\end{cases}
\end{equation}
The cavity radius $R_\mathrm{c}$ is a model parameter representing an
average distance from the atom to the nearest neighboring atoms
constituting the host medium; it has to be determined from other
(preferably microscopic) calculations or experiments. Note that
on the length scale of of the order of magnitude of $R_\mathrm{c}$,
the unperturbed host medium can be regarded as being homogeneous
from the point of view of macroscopic electrodynamics, i.e.,
\begin{equation}
\label{realcond}
\left.
\begin{array}{l}
\varepsilon(\mathbf{r},\omega)
=\varepsilon(\mathbf{r}_A,\omega)\\[.5ex]
\mu(\mathbf{r},\omega)
=\mu(\mathbf{r}_A,\omega)
\end{array}
\right\}
\ \mathrm{for}\
|\mathbf{r}-\mathbf{r}_A|\leq (1+\nu)R_\mathrm{c},
\end{equation}
with $\nu$ being a small positive number.

In the real-cavity model, the local electromagnetic field
the guest atom interacts with is the field that obeys the
macroscopic Maxwell equations, with the permittivity
$\varepsilon_\mathrm{loc}(\mathbf{r},\omega)$
and permeability $\mu_\mathrm{loc}(\mathbf{r},\omega)$
as given by Eq.~(\ref{loc}). Hence, when calculating the vdW potential
of the atom by means of Eq.~(\ref{potone}), the Green tensor therein
is obviously the Green tensor of the macroscopic Maxwell equations
with $\varepsilon_\mathrm{loc}(\mathbf{r},\omega)$ and
$\mu_\mathrm{loc}(\mathbf{r},\omega)$, respectively,  in place of
$\varepsilon(\mathbf{r},\omega)$ and $\mu(\mathbf{r},\omega)$, i.e,
the Green tensor for the electromagnetic field in the medium disturbed
by a cavity-like, small free-space inhomogeneity. Clearly, when
considering the mutual vdW interaction of two or more than two guest
atoms, each atom has to be thought of as being located at the center of
a small free-space cavity, and the Green tensors in
Eqs.~(\ref{pottwo}) and (\ref{many}) are the Green tensors of the
macroscopic Maxwell equations for the electromagnetic field in the
medium disturbed by the corresponding cavity-like, small free-space
inhomogeneities.

The task now consists in determining such a Green tensor. The
electromagnetic field inside and outside each cavity exactly solves
the macroscopic Maxwell equations, together with the standard boundary
conditions at the surface of the cavity. For an arbitrary current
distribution $\mathbf{j}(\mathbf{r},\omega)$, the electric field
$\mathbf{E}(\mathbf{r},t)$, for instance, can be given by means of the
Green tensor $\ten{G}(\mathbf{r},\mathbf{r'},\omega)$ according to
the relation
\begin{equation}
\label{electric}
\mathbf{E}(\mathbf{r},t)=i\mu_0\!\int_0^\infty
\!\mathrm{d}\omega\,\omega e^{-i\omega t}
\!\int\!\mathrm{d}^3r'\,
\ten{G}(\mathbf{r},\mathbf{r'},\omega)\!\cdot\!
\mathbf{j}(\mathbf{r'},\omega)+\mathrm{c.c.}
\end{equation}

We begin with the case where a single guest atom is embedded in a host
medium of finite size and arbitrary shape, and, at a first stage, we
assume that the host medium outside the cavity which contains the
guest atom is homogeneous---an assumption that will be dropped later.
According to Eq.~(\ref{electric}), the scattering part of the Green
tensor, $\ten{G}^{(1)}(\mathbf{r}_A,\mathbf{r}_A,\omega)$, describing
the electromagnetic field reaching the point $\mathbf{r}_A$ where it
has originated can be decomposed into three parts,
\begin{align}
\label{eq15}
\ten{G}^{(1)}(\mathbf{r}_A,\mathbf{r}_A,\omega)
=&\;\ten{G}_1^{(1)}(\mathbf{r}_A,\mathbf{r}_A,\omega)
+\ten{G}_2^{(1)}(\mathbf{r}_A,\mathbf{r}_A,\omega)
\nonumber\\
&+\ten{G}_3^{(1)}(\mathbf{r}_A,\mathbf{r}_A,\omega),
\end{align}
%
\begin{figure}[t]
\begin{center}
\includegraphics[width=0.8\linewidth]{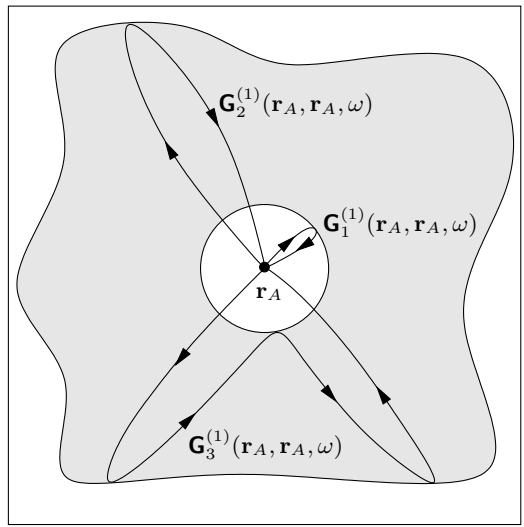}

\end{center}
\caption{Schematic illustration of the interaction of the
electromagnetic field with an atom at the center of an
empty spherical cavity (at $\mathbf{r}_A$), embedded in a
(isotropic) host medium of arbitrary size and shape. Simple
examples of the three different kinds of processes are sketched.
The corresponding Green tensors are
$\ten{G}_1^{(1)}(\mathbf{r}_A,\mathbf{r}_A,\omega)$,
accounting for all scattering processes within the cavity,
$\ten{G}_2^{(1)}(\mathbf{r}_A,\mathbf{r}_A,\omega)$, describing
(multiple) transmission through the cavity surface and accounting
for all scattering processes in the host medium excluding reflection
at the outer boundary of the cavity, and
$\ten{G}_3^{(1)}(\mathbf{r}_A,\mathbf{r}_A,\omega)$, including only
the scattering processes involving (multiple) reflection at the outer
boundary of the cavity.}
\label{single}
\end{figure}%
where $\ten{G}_1^{(1)}(\mathbf{r}_A,\mathbf{r}_A,\omega)$
accounts for (multiple) scattering at the inner surface of
the cavity, $\ten{G}_2^{(1)}(\mathbf{r}_A,\mathbf{r}_A,\omega)$
describes (multiple) transmission through the surface of the cavity
without scattering at the outer surface of the cavity, and
$\ten{G}_3^{(1)}(\mathbf{r}_A,\mathbf{r}_A,\omega)$ is associated with
(multiple) transmission through the surface of the cavity via
scattering at the outer surface of the cavity. The decomposition is
schematically illustrated in Fig.~\ref{single}, showing a simple
example each.

The term $\ten{G}_1^{(1)}(\mathbf{r}_A,\mathbf{r}_A,\omega)$ in
Eq.~(\ref{eq15}) is nothing but the scattering part of the Green
tensor at the center of an empty sphere surrounded by an infinitely
extended medium. In the case of a magnetoelectric medium of
permittivity $\varepsilon(\omega)$ and permeability $\mu(\omega)$, it
reads \cite{Li1994}
\begin{equation}
\label{greencavity}
\ten{G}_1^{(1)}(\mathbf{r}_A,\mathbf{r}_A,\omega)
=\frac{i\omega}{6\pi c}\,C(\omega)\ten{I},
\end{equation}
where
\begin{equation}
\label{greencavity1} C(\omega)
=\frac{h_1^{(1)}(z_0)\left[zh_1^{(1)}(z)\right]'
-\varepsilon(\omega) h_1^{(1)}(z)\left[z_0 h_1^{(1)}(z_0)\right]'}
{\varepsilon(\omega) h_1^{(1)}(z)\left[z_0
j_1(z_0)\right]'-j_1(z_0) \left[zh_1^{(1)}(z)\right]'}
\end{equation}
[$z_0$ $\!=$ $\!\omega R_\mathrm{c}/c$, $z$ $\!=$ $\!n(\omega) z_0$,
$n(\omega)$ $\!=$ $\!\sqrt{\varepsilon(\omega)\mu(\omega)}\,$, the
primes indicate derivatives], with $j_1(x)$ and $h_1^{(1)}(x)$,
respectively, being the first spherical Bessel function and the first
spherical Hankel function of the first kind,
\begin{gather}
j_1(x)=\frac{\sin(x)}{x^2}-\frac{\cos(x)}{x}\,,\\
h_1^{(1)}(x)=-\left(\frac{1}{x}+\frac{i}{x^2}\right)e^{ix}.
\end{gather}
At this point it should already be mentioned that since
the vdW potential depends on the Green tensor at all
frequencies, an expansion for small $|\omega|R_\mathrm{c}/c$ of the
form
\begin{multline}
\label{Cexp}
C(\omega)=3\,\frac{\varepsilon(\omega)-1}{2\varepsilon(\omega)+1}\,
\frac{c^3}{i\omega^3R_\mathrm{c}^3}\\
+\frac{9}{5}\,\frac{\varepsilon^2(\omega)[5\mu(\omega)-1]-
3\,\varepsilon(\omega)-1}{[2\varepsilon(\omega)+1]^2}\,
\frac{c}{i\omega R_\mathrm{c}}\\
+9\,\frac{\varepsilon(\omega)n^3(\omega)}{[2\varepsilon(\omega)+1]^2}
-1+O\left(\frac{\omega R_\mathrm{c}}{c}\right)
\end{multline}
and neglect of the term $O(\omega R_\mathrm{c}/c)$, as it is done in
Ref.~\cite{Ho2006} for the calculation of the decay rate of an excited
atom, has to be applied with care (see Sec.~\ref{subsub_nonlin}).

Following the arguments given in Ref.~\cite{Ho2006}, the term
$\ten{G}_2^{(1)}(\mathbf{r}_A,\mathbf{r}_A,\omega)$ in
Eq.~(\ref{eq15}) can always be written in the form
\begin{equation}
\label{combined}
\ten{G}_2^{(1)}(\mathbf{r}_A,\mathbf{r}_A,\omega)
=f^2(\omega)
\ten{G}_\mathrm{med}^{(1)}(\mathbf{r}_A,\mathbf{r}_A,\omega),
\end{equation}
where $\ten{G}_\mathrm{med}^{(1)}(\mathbf{r}_A,\mathbf{r}_A,\omega)$
is the scattering part of the Green tensor of the undisturbed host
medium (i.e., the medium without the cavity), and the global factor
$f(\omega)$ can simply be determined from examining the Green
tensor $\ten{G}_\mathrm{trans}(\mathbf{r},\mathbf{r}_A,\omega)$ which
links a source at the center of the cavity with the transmitted field
at a point $\mathbf{r}$ in the medium outside the cavity in the case
of an infinitely extended medium. For a magnetoelectric medium, the
Green tensor $\ten{G}_\mathrm{trans} (\mathbf{r},\mathbf{r}_A,\omega)$
reads \cite{Li1994}
\begin{multline}
\label{greenout}
\ten{G}_\mathrm{trans}(\mathbf{r},\mathbf{r}_A, \omega)
=\frac{\mu(\omega)k}{4\pi}\,
D(\omega)\\
\times\left[a(q)\ten{I}-b(q)\mathbf{v}\mathbf{v}\right] e^{iq}
\end{multline}
[$k$ $\!=$ $\!n(\omega)\omega/c$, $a(q)$ $\!=$ $\!1/q$ $\!+$
$\!i/q^2$ $\!-$ $\!1/q^3$, $b(q)$
$\!=$ $\!1/q$ $\!+$ $\!3i/q^2$ $\!-$ $\!3/q^3$, $q$ $\!=$
$\![n(\omega)\omega/c]|\mathbf{r}$ $\!-$ $\!\mathbf{r}_A|$,
$\mathbf{v}$ $\!=$ $\!(\mathbf{r}$ $\!-$ $\!\mathbf{r}_A)
/(|\mathbf{r}$ $\!-$ $\!\mathbf{r}_A|)$], where
\begin{multline}
\label{koeff}
D(\omega)=\\
\frac{j_1(z_0)\left[z_0h_1^{(1)}(z_0)\right]'
-\left[z_0j_1(z_0)\right]'h_1^{(1)}(z_0)}
{\mu(\omega)\Bigl\{j_1(z_0)\left[z h_1^{(1)}(z)\right]'
-\varepsilon(\omega)
\left[z_0j_1(z_0)\right]'h_1^{(1)}(z)
\Bigr\}}.
\end{multline}
Comparing Eq.~(\ref{greenout}) with the Green tensor for an
undisturbed, infinitely extended host medium,
\begin{equation}
\label{Greenbulk}
\ten{G}_\mathrm{med}^{(0)}(\mathbf{r},\mathbf{r}_A,\omega)
=\frac{\mu(\omega)k}{4\pi}
\left[a(q)\ten{I}-b(q)\mathbf{v}\mathbf{v}\right] e^{iq},
\end{equation}
we see that the factor $f(\omega)$ is equal to $D(\omega)$,
\begin{equation}
\label{eq24}
f(\omega) = D(\omega).
\end{equation}
Inspection of Eq. (\ref{koeff}) for small $|\omega|
R_\mathrm{c}/c$ shows that
\begin{equation}
\label{D}
D(\omega)=\frac{3\varepsilon(\omega)}{2\varepsilon(\omega)+1}+
O\left(\frac{\omega R_\mathrm{c}}{c}\right).
\end{equation}
Note that in leading order, the factor $D(\omega)$ is the same as that
found for a purely electric system (cf.~Ref.~\cite{Ho2006}).

As shown in Ref.~\cite{Ho2006}, the term
$\ten{G}_3^{(1)}(\mathbf{r}_A,\mathbf{r}_A,\omega)$ in
Eq.~(\ref{eq15}) behaves like $(\omega R_\mathrm{c}/c)^3$ for a
sufficiently small cavity in a dielectric medium. The same asymptotic
behavior is observed for magnetoelectric media, since, for
sufficiently small $R_\mathrm{c}$, the leading-order contribution to
$\ten{G}_3^{(1)}(\mathbf{r}_A,\mathbf{r}_A,\omega)$ is determined by
the electric properties. Recalling that in the real-cavity model the
cavity radius is assumed to be small compared with the relevant atomic
and medium wavelengths,
$\ten{G}_3^{(1)}(\mathbf{r}_A,\mathbf{r}_A,\omega)$ is in any case
small compared with
$\ten{G}_1^{(1)}(\mathbf{r}_A,\mathbf{r}_A,\omega)$ and
$\ten{G}_2^{(1)}(\mathbf{r}_A,\mathbf{r}_A,\omega)$ and can thus be
neglected, so that Eq.~(\ref{eq15}) [together with
Eqs.~(\ref{greencavity}), (\ref{combined}), and (\ref{eq24})] takes
the form
\begin{equation}
\label{greenzerleg}
\ten{G}^{(1)}(\mathbf{r}_A,\mathbf{r}_A,\omega)
=\frac{i\omega}{6\pi}\,C(\omega)\ten{I}
+D^2(\omega)
\ten{G}^{(1)}_\mathrm{med}(\mathbf{r}_A,\mathbf{r}_A, \omega).
\end{equation}

So far we have concentrated on homogeneous host media of arbitrary
shapes. An inhomogeneous medium can always be divided into a
(small) homogeneous part that contains the cavity, such that the
condition (\ref{realcond}) is satisfied, and an inhomogeneous
rest. Since all the scattering processes resulting from the
inhomogeneous rest can be thought of as being taken into account
by the scattering part of the Green tensor of the host medium,
Eq.~(\ref{greenzerleg}) still holds in this more general case.
Note that a homogeneous medium of finite size can already be
regarded as a special case of an inhomogeneous system. Hence,
$\varepsilon(\omega)$ and $\mu(\omega)$, respectively, can be
safely replaced by $\varepsilon_A(\omega)$ $\!\equiv$
$\!\varepsilon(\mathbf{r}_A,\omega)$ and $\mu_A(\omega)$
$\!\equiv$ $\!\mu(\mathbf{r}_A,\omega)$ in the equations given
above [$C(\omega)\mapsto C_A(\omega)$, $D(\omega)\mapsto
D_A(\omega)$].

\begin{figure}[t]
\begin{center}
\includegraphics[width=0.8\linewidth]{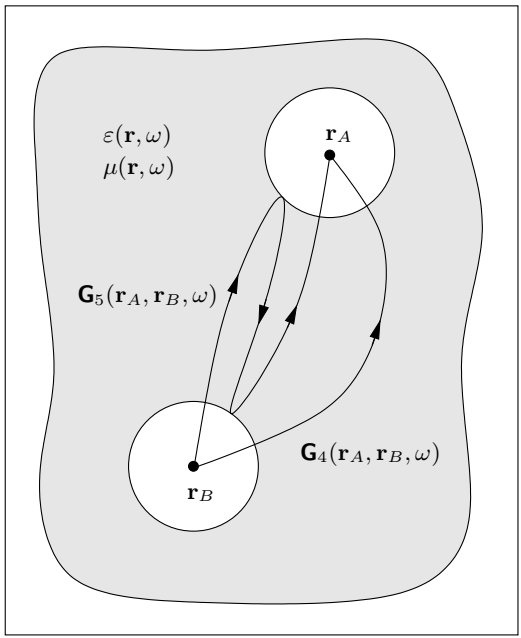}
\end{center}
\caption{Schematic illustration of the interaction of the
electromagnetic field with two atoms, each placed in a
spherical cavity, surrounded by a host medium of arbitrary size
and shape. The relevant Green tensors
$\ten{G}_4(\mathbf{r}_A,\mathbf{r}_A,\omega)$ and
$\ten{G}_5(\mathbf{r}_A,\mathbf{r}_A,\omega)$ account for
(multiple) transmission through the cavity surfaces and
(multiple) scattering at the inhomogeneities of the host medium
including (multiple) scattering at the outer boundaries of the
cavities, respectively.
}
\label{two}
\end{figure}

Next we consider two atoms within a host medium of arbitrary size
and shape, both located at the center of a small cavity, such that
Eqs. (\ref{loc}) and (\ref{realcond}) hold. In this case, there are
two contributions to the Green tensor
$\ten{G}(\mathbf{r}_A,\mathbf{r}_B,\omega)$ that determines the
electromagnetic field at the position $\mathbf{r}_A$ of atom $A$,
originating at $\mathbf{r}_B$ where is atom~$B$ located,
\begin{equation}
\ten{G}(\mathbf{r}_A,\mathbf{r}_B,\omega)=
\ten{G}_4(\mathbf{r}_A,\mathbf{r}_B,\omega)+
\ten{G}_5(\mathbf{r}_A,\mathbf{r}_B,\omega).
\end{equation}
Here, $\ten{G}_4(\mathbf{r}_A,\mathbf{r}_B,\omega)$ accounts for
(multiple) transmission through the cavity surfaces and (multiple)
scattering at the inhomogeneities of the medium, excluding scattering
at the outer boundaries of the cavities. As in the single-atom case,
processes that involve (multiple) scattering at the outer surfaces of
the cavities, described by
$\ten{G}_5(\mathbf{r}_A,\mathbf{r}_B,\omega)$, are
neglected. The situation is sketched in Fig.~\ref{two}.

By means of the same arguments as in the single-atom case, we find a
global factor that accounts for transmission through the surfaces
of the cavities such that the field at $\mathbf{r}$ outside the
cavities which results from $\mathbf{r}_B$, may be calculated from
$D_B(\omega) \ten{G}_\mathrm{med}(\mathbf{r},\mathbf{r}_B,\omega)$.
Due to the property $\ten{G}(\mathbf{r},\mathbf{r'},\omega)$ $\!=$
$\!\ten{G}^\mathsf{T}(\mathbf{r'},\mathbf{r},\omega)$
of the Green tensor, we also find that
$D_A(\omega)\ten{G}_\mathrm{med}(\mathbf{r}_A,\mathbf{r},\omega)$
determines the electromagnetic field at point $\mathbf{r}_A$
originating at point $\mathbf{r}$ outside the cavities. We may thus
write $\ten{G}(\mathbf{r}_A,\mathbf{r}_B,\omega)$ in the form
\begin{equation}
\label{greenzerleg3}
\ten{G}(\mathbf{r}_A,\mathbf{r}_B,\omega)=
D_A(\omega)
\ten{G}_\mathrm{med}(\mathbf{r}_A,\mathbf{r}_B,\omega)D_B(\omega).
\end{equation}


\section{Local-field corrected van der Waals interaction}
\label{sec_local}


\subsection{Single atom within an arbitrary magnetoelectric medium}
\label{sub_one}

Within the framework of the real-cavity model, local-field
corrected vdW interactions can now be calculated by combining
the formulas given in Sec.~\ref{sec_basic} and \ref{sec_real}, where
we again begin with the case of a single guest atom. We substitute
$\ten{G}{^{(1)}}(\mathbf{r}_A,\mathbf{r}_A,\omega)$ as given by
Eq.~(\ref{greenzerleg}) into Eq.~(\ref{potone}) to obtain
\begin{equation}
\label{Utotal}
U(\mathbf{r}_A)=U_1(\mathbf{r}_A)+U_2(\mathbf{r}_A),
\end{equation}
where
\begin{align}
\label{U1}
U_1(\mathbf{r}_A)=&\,-\frac{\hbar\mu_0}{12\pi^2 c}
\int _0^{\infty}\mathrm{d} u\, u^3 C_A(iu)
\mathrm{Tr}\,\bm{\alpha}_A(iu),
\\
\label{U2}
U_2(\mathbf{r}_A)=&\,\frac{\hbar\mu_0}{2\pi}
\int _0^{\infty}\mathrm{d} u\, u^2
D_A^2(iu)
\nonumber\\  &\times\mathrm{Tr}\!\left[
\bm{\alpha}_A(iu)\!\cdot\!
\ten{G}_\mathrm{med}^{(1)}(\mathbf{r}_A,\mathbf{r}_A,iu)\right]
\end{align}
[$C_A(iu)$ and $D_A(iu)$ according to Eqs.~(\ref{greencavity1})
and (\ref{koeff}), respectively].


\subsubsection{Linear approximation}
\label{subsub_linear}

In order to make contact with a microscopic description and provide a
deeper insight into the validity of the real-cavity model, let us
first restrict our attention to to weakly polarizable and/or
magnetizable host material so that the linear approximation applies.
In this case, we may regard $\varepsilon(\mathbf{r},\omega)$ and
$\mu(\mathbf{r},\omega)$ as being only slightly perturbed from their
free-space values,
\begin{align}
\varepsilon(\mathbf{r},\omega)=&\,1+\chi(\mathbf{r},\omega),
\\ \mu(\mathbf{r},\omega)=&\,1+\zeta(\mathbf{r},\omega)
\end{align}
[$|\chi(\mathbf{r},\omega)|$ $\!\ll$ $\!1$,
$|\zeta(\mathbf{r},\omega)|$ $\!\ll$ $\!1$],
and approximate $U_1(\mathbf{r}_A)$, Eq.~(\ref{U1}), by
\begin{multline}
U_1(\mathbf{r}_A)=-\frac{\hbar\mu_0}{4\pi^2c}
\int _0 ^\infty \mathrm{d} u \,u^3 \alpha_A(iu)e^{-2uR_\mathrm{c}/c}
\\
\times \left[\left(\frac{c^3}{u^3R_\mathrm{c}^3}
+\frac{2c^2}{u^2R_\mathrm{c}^2}+\frac{c}{uR_\mathrm{c}}
+\frac{1}{2}\right)\chi(\mathbf{r}_A,iu)\right.
\\
\left.-\left(\frac{c}{uR_\mathrm{c}}
+\frac{1}{2}\right)\zeta(\mathbf{r}_A,iu)\right],
\end{multline}
where we have assumed a spherically symmetric atom.
By straightforward calculation, one can then show that
\begin{equation}
\label{bornU1}
U_1(\mathbf{r}_A)=\rho(\mathbf{r}_A)
\int _{|\mathbf{s}-\mathbf{r}_A|\geq R_\mathbf{c}}
\mathrm{d}^3s\,
U(\mathbf{r}_A,\mathbf{s}),
\end{equation}
where $\rho(\mathbf{r})$ is the number density of the medium atoms,
\begin{equation}
\label{U2atom}
U(\mathbf{r}_A,\mathbf{s})=U^\mathrm{el}(\mathbf{r}_A,\mathbf{s})
+U^\mathrm{mag}(\mathbf{r}_A,\mathbf{s})
\end{equation}
($\mathbf{r}_A$ $\!\neq$ $\!\mathbf{s}$) is the vdW potential of two
unpolarized, neutral (ground state) atoms in free space, with the
electric and magnetic parts being given by \cite{review2006}
\begin{multline}
\label{U2atom_el}
U^\mathrm{el}(\mathbf{r}_A,\mathbf{r})
=-\frac{\hbar\mu_0^2}{2\pi}
\int _0^\infty \mathrm{d} u\, u^4\alpha_A(iu)\alpha(iu)\\
\times \mathrm{Tr}[\ten{G}_\mathrm{free}
(\mathbf{r}_A,\mathbf{r},iu)\!\cdot\!
\ten{G}_\mathrm{free}(\mathbf{r},\mathbf{r}_A,iu)]\\
=-\frac{\hbar}{32\pi^3\varepsilon_0^2 l^6}
\int _0^\infty \mathrm{d}u\, \alpha_A(iu)\alpha(iu)g(ul/c)
\end{multline}
and
\begin{multline}
\label{U2atom_mag}
U^\mathrm{mag}(\mathbf{r}_A,\mathbf{r})
=-\frac{\hbar\mu_0^2}{2\pi}
\int _0^\infty \mathrm{d}u\, u^2 \alpha_A(iu) \beta(iu)
\\
\times\mathrm{Tr}\!\left\{\left[
\ten{G}_\mathrm{free}(\mathbf{r}_A,\mathbf{s},iu)
\!\times\!\overleftarrow{\bm{\nabla}}_\mathbf{s}\right]
\!\cdot\!\bm{\nabla}_\mathbf{s}\!\times\!
\ten{G}_\mathrm{free}(\mathbf{s},\mathbf{r}_A,iu)\right\}
_{\mathbf{s}=\mathbf{r}}
\\
=\frac{\hbar\mu_0^2 }{32\pi^3l^4} \int _0^\infty \mathrm{d}u\,
u^2\alpha_A(iu)\beta(iu)h(ul/c)
\end{multline}
[$g(x)=2e^{-2x}(3+6x+5x^2+2x^3+x^4)$, $h(x)=2e^{-2x}(1+2x+x^2)$,
$l=|\mathbf{r}_A-\mathbf{r}|$;
$\ten{G}_\mathrm{free}(\mathbf{r},\mathbf{s},iu)$, free-space Green
tensor; $\alpha(iu)$, polarizability; $\beta(iu)$, magnetizability
of the medium atoms], where we have made the identifications
\begin{align}
\label{chi}
&\chi(\mathbf{r},\omega)
=\varepsilon_0^{-1}\rho(\mathbf{r})\alpha(\omega),\\
\label{zeta}
&\zeta(\mathbf{r},\omega)=\mu_0\rho(\mathbf{r})\beta(iu).
\end{align}

To find the linear approximation to $U_2(\mathbf{r}_A)$,
Eq.~(\ref{U2}), let us first consider
\begin{align}
\label{eq41}
\ten{G}_\mathrm{med}^{(1)}(\mathbf{r}_A,\mathbf{r}_A,\omega)
=\ten{G}_\mathrm{med}(\mathbf{r}_A,\mathbf{r}_A,\omega)-
\ten{G}_\mathrm{med}^{(0)}(\mathbf{r}_A,\mathbf{r}_A,\omega).
\end{align}
Applying the linear Born expansion (see, e.g., Ref.~\cite{Ho2006})
\begin{align}
\label{born1}
&\ten{G}(\mathbf{r},\mathbf{r}',\omega)
=\ten{G}_\mathrm{free}(\mathbf{r},\mathbf{r}',\omega)+
\Delta\ten{G}(\mathbf{r},\mathbf{r}',\omega)\\
\label{born2}
&\Delta \ten{G}(\mathbf{r},\mathbf{r}',\omega)
=\Delta \ten{G}^{\mathrm{el}}(\mathbf{r},\mathbf{r}',\omega)
+\Delta\ten{G}^{\mathrm{mag}}(\mathbf{r},\mathbf{r}',\omega),
\end{align}
with
\begin{multline}
\label{born3}
\Delta\ten{G}^{\mathrm{el}}(\mathbf{r},\mathbf{r}',\omega)
=\frac{\omega^2}{c^2}
\int \mathrm{d}^3s\, \chi(\mathbf{s},\omega)\\
\times\ten{G}_\mathrm{free}(\mathbf{r},\mathbf{s},\omega)
\!\cdot\!
\ten{G}_\mathrm{free}(\mathbf{s},\mathbf{r}',\omega),
\end{multline}
\begin{multline}
\label{born4}
\Delta\ten{G}^\mathrm{mag}(\mathbf{r},\mathbf{r}',\omega)
=-\int \mathrm{d}^3s\, \zeta(\mathbf{s},\omega)
\\
\times\left[\ten{G}_\mathrm{free}(\mathbf{r},\mathbf{s},\omega)
\!\times\!\overleftarrow{\bm{\nabla}}_s\right]\!\cdot\!\bm{\nabla}_s
\!\times\!\ten{G}_\mathrm{free}(\mathbf{s},\mathbf{r}',\omega)
\end{multline}
to the calculation of the Green functions
$\ten{G}_\mathrm{med}(\mathbf{r}_A,\mathbf{r}_A,\omega)$
and $\ten{G}_\mathrm{med}^{(0)}(\mathbf{r}_A,\mathbf{r}_A,\omega)$
in Eq.~(\ref{eq41}), we derive
\begin{multline}
\label{greenborn}
\ten{G}^{(1)}_\mathrm{med}(\mathbf{r}_A,\mathbf{r}_A,\omega) =
\\
\frac{\omega^2}{c^2} \int _\mathrm{V} \mathrm{d}^3s\,
\chi(\mathbf{s},\omega)
\ten{G}_\mathrm{free}(\mathbf{r}_A,\mathbf{s},\omega)\!\cdot\!
\ten{G}_\mathrm{free}(\mathbf{s},\mathbf{r}_A,\omega)
\\
-\int _\mathrm{V}\mathrm{d}^3s\, \zeta(\mathbf{s},\omega)
\left[\ten{G}_\mathrm{free}(\mathbf{r},\mathbf{s},\omega)
\!\times\!\overleftarrow{\bm{\nabla}}_s\right]\!\cdot\!\bm{\nabla}_s
\!\times\!\ten{G}_\mathrm{free}(\mathbf{s},\mathbf{r}',\omega)
\\
-\frac{\omega^2\chi(\mathbf{r}_A,\omega) }{c^2} \int
\mathrm{d}^3s\,
\ten{G}_\mathrm{free}(\mathbf{r}_A,\mathbf{s},\omega)\!\cdot\!
\ten{G}_\mathrm{free}(\mathbf{s},\mathbf{r}_A,\omega)\\
+\,\zeta(\mathbf{r}_A,\omega)
\int \mathrm{d}^3s
\left[\ten{G}_\mathrm{free}(\mathbf{r},\mathbf{s},\omega)
\!\times\!\overleftarrow{\bm{\nabla}}_s\right]\!\cdot\!\bm{\nabla}_s
\!\times\!\ten{G}_\mathrm{free}(\mathbf{s},\mathbf{r}',\omega)
\end{multline}
($V$, volume of the unperturbed host medium). Equation
(\ref{greenborn}) reveals that, in contrast to $D_A(\omega)$,
$\ten{G}_\mathrm{med}^{(1)}(\mathbf{r}_A,\mathbf{r}_A,\omega)$
has no zero-order contribution with respect to
$\chi(\mathbf{r},\omega)$ and $\zeta(\mathbf{r},\omega)$. Thus, the
linear approximation to $U_2(\mathbf{r}_A)$ can be obtained by
substituting Eq.~(\ref{greenborn}) into Eq.~(\ref{U2}), letting
$D_A(\omega)$ $\!=$ $\!1$, and applying Eqs.~(\ref{chi}) and
(\ref{zeta}), which results in
\begin{align}
\label{U2born}
U_2(\mathbf{r}_A)=
&\int _\mathrm{V}\mathrm{d}^3s\,
\rho(\mathbf{s}) U(\mathbf{r}_A,\mathbf{s})
\nonumber\\
&\;-\rho(\mathbf{r}_A)
\int \mathrm{d}^3s\,
U(\mathbf{r}_A,\mathbf{s}).
\end{align}
As expected, the overall potential in linear approximation is simply
\begin{align}
\label{eq44}
U(\mathbf{r}_A)& = U_1(\mathbf{r}_A)+U_2(\mathbf{r}_A)
\nonumber\\
&= \int _\mathrm{V'} \mathrm{d}^3s\,
\rho(\mathbf{s})U(\mathbf{r}_A,\mathbf{s})
\end{align}
($\mathrm{V'}$, volume of the host medium excluding the volume of
the cavity), where we have recalled Eq.~(\ref{realcond}).

The results found in linear approximation, particularly
Eq.~(\ref{eq44}), indicate that $R_\mathrm{c}$ can indeed be
identified with an average inter-atomic distance. Hence, the
real-cavity model can be regarded as being valid down to a microscopic
scale. Since $U_1(\mathbf{r}_A)$ can be described as the sum of the
vdW interactions between the guest atom and all the atoms in an
infinitely extended homogeneous medium, it does not lead to a force
but to a more or less significant shift of the overall potential,
which is dominated by the interaction of the guest atom with the
neighboring host atoms. Note that we have demonstrated these results
up to linear order in $\chi(\mathbf{r},\omega)$ and
$\zeta(\mathbf{r},\omega)$. In principle, a generalization
beyond the linear Born expansion should be possible, although
associated with extensive calculations. In this case many-atom vdW
interactions have to be taken into account.


\subsubsection{Beyond the linear approximation}
\label{subsub_nonlin}

Let us return to the integrals in Eqs.~(\ref{U1}) and (\ref{U2}).
To find the local-field correction, one has in fact to consider
the leading-order terms in $R_\mathrm{c}$ after the frequency
integrals have been performed. However, since
$\varepsilon_A(\omega)-1$, $\mu_A(\omega)-1$, and
$\alpha_A(\omega)$ provide cut-off functions for the $u$-integrals
[giving their main contribution for $u\lesssim\omega_{\max}$, with
$\omega_{\max}$ being the maximum of all relevant atomic and
medium resonance frequencies] and assuming that
$\omega_{\max}R_\mathrm{c}/c\ll 1$, we may expand $C_A(iu)$ and
$D_A(iu)$ in the integrands, keep the leading-order terms in
$uR_\mathrm{c}/c$, and integrate over $u$ afterwards. This procedure
can be justified by the good convergence behavior in the asymptotic
limit of small $R_\mathrm{c}$, which we have confirmed numerically.

Thus, using Eq.~(\ref{Cexp}), we can write $U_1(\mathbf{r}_A)$,
Eq.~(\ref{U1}), in the form
\begin{multline}
\label{UCapprox}
U_1(\mathbf{r}_A)=-\frac{\hbar\mu_0}{4\pi^2}
\int_0^{\infty}\mathrm{d} u
\left[\frac{\varepsilon_A(iu)-1}{2\varepsilon_A(iu)+1}\,
\frac{c^2}{R_\mathrm{c}^3}\right.\\
\left. +\,3\,\frac{\varepsilon_A^2(iu)\left[1-5\mu_A(iu)\right]+
3\varepsilon_A(iu)+1}{5\left[2\varepsilon_A(iu)+1\right]^2}\,
\frac{u^2 }{R_\mathrm{c}}\right]\\
\times\mathrm{Tr}\,\bm{\alpha}_A(iu).
\end{multline}
As already mentioned, $U_1(\mathbf{r}_A)$ can be ignored if one is
only interested in the force on the atom (see also App.~\ref{app1}).
Similarly, using Eq.~(\ref{D}), we can write $U_2(\mathbf{r}_A)$,
Eq.~(\ref{U2}), as
\begin{multline}
\label{U2approx}
U_2(\mathbf{r}_A)=\frac{\hbar \mu_0}{2\pi} \int _0 ^\infty
\mathrm{d}u u^2
\left[\frac{3\varepsilon_A(iu)}{2\varepsilon_A(iu)+1}\right]^2\\
\times\mathrm{Tr}\left[\bm{\alpha}_A(iu)\!\cdot\!
\ten{G}_\mathrm{med}^{(1)}(\mathbf{r}_A,\mathbf{r}_A,iu)\right].
\end{multline}
With Eq.~(\ref{forceone}) and the condition (\ref{realcond}), one
obtains
\begin{align}
\label{forceloc}
\mathbf{F}(\mathbf{r}_A)=&-\frac{\hbar\mu_0}{2\pi}
\int_0^{\infty}\mathrm{d}u u^2
\left[\frac{3\varepsilon_A(iu)}{2\varepsilon_A(iu)+1}\right]^2\nonumber\\
&\times\bm{\nabla}_{\!\!A}
\mathrm{Tr}\left[\bm{\alpha}_A(iu)\cdot
\ten{G}_\mathrm{med}^{(1)}(\mathbf{r}_A,\mathbf{r}_A,iu)\right]
\end{align}
for the local-field corrected vdW force on a ground-state atom
within an arbitrary magnetoelectric system.

In the limit of the atom being embedded in a very dilute medium,
$\varepsilon_A(iu)-1$ $\!\ll$ $\!1$ and $\mu_A(iu)-1$ $\!\ll$ $\!1$,
the local-field correction becomes negligible and
Eqs.~(\ref{UCapprox})--(\ref{forceloc}) approach the uncorrected
result [Eqs.~(\ref{potone}) and (\ref{forceone})], as expected. As a
note of caution, we remark that, quite contrary, in the limit
$R_\mathrm{c}$ $\!\to$ $\!0$, the local-field corrected result does
not approach the uncorrected one, which might be obtained when the
atom is thought of as not being placed in an empty free-space region,
but being placed in a region which is continuously filled with medium.
This is due to the fact that there is no continuous $R_\mathrm{c}$
$\!\to$ $\!0$ transition from the corrected setup (an atom inside an
empty free-space cavity surrounded by a medium) and the uncorrected
one (an atom inside the medium): Even as the cavity radius is made
smaller and smaller, the atom will always remain inside a free-space
region.

Equation~(\ref{forceloc}) shows that, in comparison to the uncorrected
vdW force~(\ref{forceone}), the contributions to the
\mbox{$u$-in\-tegral} are enhanced by a factor between $1$ and $9/4$
owing to the local-field correction. Note that the enhancement factor
varies with $u$ and the uncorrected contributions can in general (in
particular, for magnetoelectric media) have different signs for
different values of $u$. Hence, the local-field corrected vdW force
can be enhanced or reduced. For example, for (purely electric) media
important in biological applications, local-field corrections can lead
to enhancements by factors of $2.22$ [for water with
$\varepsilon(0)\approx 80$] or $1.86\ldots 2.19$ (for typical plasma
membrane compositions with $\varepsilon(0)\approx 5 \ldots 40$
\cite{Nir1976}).


\subsection{Two or more atoms within an arbitrary magnetoelectric
medium}
\label{subsec_two}

Using Eqs.~(\ref{pottwo}) and (\ref{greenzerleg3}), the local-field
corrected two-atom vdW potential reads
\begin{multline}
\label{U_AB}
U(\mathbf{r}_A,\mathbf{r}_B)=-\frac{\hbar\mu_0^2}{2\pi}
\int _0^{\infty}\mathrm{d}u u^4 D_A(iu)D_B(iu)\\
\times\mathrm{Tr}[\bm{\alpha}_A(iu)\!\cdot\!
\ten{G}_\mathrm{med}(\mathbf{r}_A,\mathbf{r}_B,iu)
\!\cdot\!\bm{\alpha}_B(iu)\!\cdot\!
\ten{G}_\mathrm{med}(\mathbf{r}_B,\mathbf{r}_A,iu)].
\\[-2ex]
\end{multline}
In analogy to Sec.~\ref{subsub_nonlin}, we may expand the
coefficients $D_{A}(iu)$ and $D_B(iu)$ to leading order in
$uR_\mathrm{c}/c$, as given by Eq.~(\ref{D}), and integrate
afterwards. Hence, Eq.~(\ref{U_AB}) leads to
\begin{multline}
\label{UABl}
U(\mathbf{r_A},\mathbf{r_B})=-\frac{\hbar\mu_0^2}{2\pi}
\\\times
\int_0^{\infty}\mathrm{d}u u^4
\left[\frac{3\varepsilon_A(iu)}{2\varepsilon_A(iu)+1}\right]^2
\left[\frac{3\varepsilon_B(iu)}{2\varepsilon_B(iu)+1}\right]^2
\\\times\!
\mathrm{Tr}[\bm{\alpha}_A(iu)\!\cdot\!
\ten{G}_\mathrm{med}(\mathbf{r}_A,\mathbf{r}_B,iu)
\!\cdot\!\bm{\alpha}_B(iu)\!\cdot\!
\ten{G}_\mathrm{med}(\mathbf{r}_B,\mathbf{r}_A,iu)].
\\[-2ex]
\end{multline}
In comparison to the uncorrected potential~(\ref{pottwo}), the
contributions to the $u$-integral are enhanced by a factor which
varies between $1$ and $81/16$, so that altogether an enhancement
or an reduction of the local-field corrected two-atom force may be
observed (recall the remark at the end of Sec.~\ref{subsub_nonlin}).

Equation~(\ref{UABl}) can easily be extended to the mutual
interaction of $N$ atoms. Substituting Eqs.~(\ref{D}) and
(\ref{greenzerleg3}) into Eq.~(\ref{many}), one finds that,
for isotropic atoms, the local-field corrected $N$-atom potential is
given by
\begin{multline}
\label{UN}
U(\mathbf{r}_1,\dots,\mathbf{r}_N)=\frac{(-1)^{N-1}\hbar\mu_0^N}
{(1+\delta_{2N})\pi}\int _0^\infty \mathrm{d}u\,u^{2N}\\
\times
\left[\frac{3\varepsilon_1(iu)}{2\varepsilon_1(iu)+1}\right]^2\cdots
\left[\frac{3\varepsilon_N(iu)}{2\varepsilon_N(iu)+1}\right]^2
\alpha_1(iu)\cdots\alpha_N(iu)\\
\times
\mathrm{Tr}\left[\ten{G}_\mathrm{med}(\mathbf{r}_1,\mathbf{r}_2,iu)
\cdots\ten{G}_\mathrm{med}(\mathbf{r}_N,\mathbf{r}_1,iu)\right].
\end{multline}


\subsection{Example: Two atoms within a bulk magnetoelectric medium}
\label{subsec_bulk}

As a simple example, let us consider two isotropic ground-state atoms
embedded in magnetoelectric bulk material characterized by
$\varepsilon(\omega)$ and $\mu(\omega)$. Substituting
Eq.~(\ref{Greenbulk}) into Eq.~(\ref{UABl}), we obtain
\begin{multline}
\label{twobulk}
U(\mathbf{r}_A,\mathbf{r}_B)=
-\frac{\hbar}{16\pi^3\varepsilon^2_0 l^6 } \int _0^{\infty}
\mathrm{d}u\,\frac{\alpha_A(iu)\alpha_B(iu)}{\varepsilon^2(iu)}\\
\times\left[\frac{3\varepsilon(iu)}{2\varepsilon(iu)+1}\right]^4
\left\{3+6n(iu)ul/c+5[n(iu)ul/c]^2\right.\\
\left.+2[n(iu)ul/c]^3+[n(iu)ul/c]^4\right\}e^{-2n(iu)ul/c},
\end{multline}
where $l=$ $\!|\mathbf{r}_A$ $\!-$ $\!\mathbf{r}_B|$. In the retarded
limit, $l$ $\!\gg$ $\!c/\omega_{\min}$ (with $\omega_{\min}$ denoting
the minimum of all the relevant atomic and medium resonance
frequencies), the main contributions to the $u$-integral come from the
range where
\begin{equation}
\alpha_{A(B)}(iu)\simeq\alpha_{A(B)}(0),\; \varepsilon(iu)\simeq
\varepsilon(0), \; \mu(iu)\simeq\mu(0).
\end{equation}
With this inserted, Eq.~(\ref{twobulk}) simplifies to
\begin{equation}
\label{eq57}
U(\mathbf{r}_A,\mathbf{r}_B)=-\frac{C_\mathrm{r}}{l^7},
\end{equation}
where
\begin{equation}
C_{\mathrm{r}}=\frac{23\hbar c}{64\pi^3\varepsilon_0^2 }
\frac{\alpha_A(0)\alpha_B(0)}{n(0)\varepsilon^2(0)}
\left[\frac{3\varepsilon(0)}{2\varepsilon(0)+1}\right]^4.
\end{equation}
In the non-retarded limit, \mbox{$l$ $\!\ll$
$\!c/n(0)\omega_{\max}$}, we may set $e^{-2n(iu)ul/c}$ $\!\simeq$
$\!1$ and approximate the term in the curly brackets by $3$. In this
way, Eq.~(\ref{twobulk}) reduces to
\begin{equation}
\label{eq59}
U(\mathbf{r}_A,\mathbf{r}_B)=-\frac{C_\mathrm{nr}}{l^6},
\end{equation}
where
\begin{equation}
C_{\mathrm{nr}}=\frac{3\hbar}{16\pi^3\varepsilon_0^2}
\int_0^{\infty} \mathrm{d}u\,
\frac{\alpha_A(iu)\alpha_B(iu)}{\varepsilon^2(iu)}
\left[\frac{3\varepsilon(iu)}{2\varepsilon(iu)+1}\right]^4.
\end{equation}

Equations~(\ref{eq57}) and (\ref{eq59}) generalize the well-known
results for the uncorrected mutual vdW interaction of two atoms in
bulk material \cite{ReviewLifshitz}. We see that the local-field
corrected potential follows the usual $l^{-7}$ and $l^{-6}$ dependence
in the retarded and non-retarded limits, respectively. However, due
to the influence of the local-field correction, the
potential/force can be noticeably enhanced in
dense media.


\section{Summary}
\label{sum}

In this paper, we have studied the local-field corrected vdW
interaction of ground-state atoms embedded in a finite magnetoelectric
medium of arbitrary size and shape. We have started from the familiar
expressions for the one- and many-atom vdW potentials in leading-order
perturbation theory, given in terms of the associated (classical)
Green tensors and the atomic polarizabilities. Employing the
real-cavity model for the local-field correction and using the
methods developed in Ref.~\cite{Ho2006}, we have derived general
expressions for the required local-field corrected Green tensors.
Using these results, we have obtained geometry-independent, general
formulas for the local-field corrected one- and many-atom
(ground-state) vdW potentials and the corresponding forces.

The derivation is based on the assumption that the radius of the
free-space cavity, in which a guest atom is thought of as being
located and which is a measure of the average distance of the guest
atom from the surrounding host atoms, is small compared to the other
characteristic lengths of the problem. That is to say, it should be
small compared to the distance of a guest atom from surfaces of
discontinuity of the host medium, the distances between the guest
atoms in the case of many-atom interactions, and the wavelengths
associated with the electromagnetic response of the guest atoms
and the host medium. It should be pointed out that under these
conditions the forces do not explicitly depend on the cavity radius.

We have found that the local-field correction leads to a
position-independent shift of the single-atom potential, while at the
same time enhancing the contributions to the position-dependent part
by a frequency-dependent factor that may vary between $1$ and $9/4$.
The frequency-dependent contributions to the two-atom potential, on
the other hand, are enhanced by a factor between $1$ and $81/16$ owing
to the local-field correction. The contributions to the general
$N$-atom potential are enhanced accordingly. In all cases, the
respective vdW force can be enhanced or reduced.

As a simple example, we have calculated the mutual vdW potential
of two ground-state atoms placed within a magnetoelectric bulk medium.
The result shows that the potential follows the well-known asymptotic
$l^{-7}$ and $l^{-6}$ power laws in the retarded and non-retarded
regimes, respectively, where the local-field correction leads to an
enhancement of the associated proportionality constants.

We conclude by noting that, besides vdW interaction and spontaneous
decay, the local-field corrected Green tensors obtained can also be
used to account for the effects of the local field in a number of
important processes (e.g., fluorescence, light scattering, energy
transfer, etc.). Evidently, one can expect that these processes are
affected by local-field corrections in a similar way as found here.


\acknowledgments

This work was supported by the Deutsche Forschungsgemeinschaft and
by the Ministry of Science of the Republic of Croatia. A.S.,
S.Y.B., and D.-G.W.\ are grateful to Ho Trung Dung for
discussions.


\appendix
\section{Atom inside a free-space cavity}
\label{app1}

In Sec.~\ref{sec_local}, we have encountered the potential term
$U_1(\mathbf{r}_A)$, Eq.~(\ref{U1}), which is equal to the potential
of an atom at the center of a free-space cavity embedded in a bulk
medium of permittivity $\varepsilon_A(\omega)$ and permeability
$\mu_A(\omega)$. More generally, the potential of an atom at a small
displacement $\overline{\mathbf{r}}_A$ from the center of the cavity
is given by Eq.~(\ref{potone}) with the Green tensor \cite{Li1994}
\begin{multline}
\ten{G}^{(1)}(\mathbf{r}_A,\mathbf{r}_A,\omega)
=\frac{i\omega}{4\pi c}\sum_{l=1}^\infty
(2l+1)\Bigl\{C^E_l(\omega)l(l+1)\frac{j^2_l(\xi_A)}{\xi^2_A}\\
\times\,\mathbf{e}_{r}\mathbf{e}_{r}
+\frac{1}{2}\left[C_l^E(\omega)\frac{[\xi_Aj_l(\xi_A)]'^2}{\xi^2_A}
+C^M_l(\omega)j^2_l(\xi_A)\right]\\
\times(\mathbf{e}_{\theta}
\mathbf{e}_{\theta}+\mathbf{e}_{\phi}\mathbf{e}_{\phi})\Bigr\}.
\end{multline}
($\mathbf{e}_{r}$, $\mathbf{e}_{\theta}$, $\mathbf{e}_{\phi}$,
spherical unit vectors associated with $\overline{\mathbf{r}}_A$;
$\xi_A$ $\!=$ $\!\omega \overline{r}_A/c$). Here $C^E_l$ is given by
Eq.~(\ref{greencavity1}), with the subscript $1\rightarrow l$, and
$C^M_l$ $\!=$ $\!C^E_l[\varepsilon\leftrightarrow\mu]$. From
\begin{equation}
j_l(\xi_A)\simeq\frac{\xi^l_A}{(2l+1)!!}\quad\mbox{for }\xi_A\ll 1,
\end{equation}
we have
\begin{multline}
\ten{G}^{(1)}(\mathbf{r}_A,\mathbf{r}_A,\omega)
=\frac{i\omega}{4\pi c}\sum_{l=1}^\infty
(2l+1)C_l^E(\omega)\biggl[\frac{\xi^{l-1}_A}{(2l+1)!!}\biggr]^2\\
\times\bigl[l(l+1)\mathbf{e}_{r}\mathbf{e}_{r}+
{\textstyle\frac{1}{2}}(l+1)^2(\mathbf{e}_{\theta}\mathbf{e}_{\theta}+
\mathbf{e}_{\phi}\mathbf{e}_{\phi})\bigr]
\end{multline}
for the small $\xi_A$ expansion of the Green tensor. As seen,
in the limit $\xi_A\rightarrow 0$, only the $l$ $\!=$ $\!1$ term survives,
leading to the potential~(\ref{U1}). The force on an atom close to the
cavity center is determined by the derivative of the Green function
with respect to $\overline{r}_A$. Therefore, for small displacements
of the atom from the cavity center, it is given by the $l$ $\!=$ $\!2$
term in the above expansion, and we find the linear force
\begin{equation}
\mathbf{F}({\mathbf r}_A)=
-\mathbf{e}_{r}\frac{\partial}{\partial\overline{r}_A}\,
U({\mathbf r}_A)
=K\overline{\mathbf{r}}_A
\end{equation}
with
\begin{equation}
\label{K}
K=-\frac{\hbar}{12\pi^2\varepsilon_0c^5}
\int_0^\infty\mathrm{d}u\,u^5\alpha_A(iu)C_2^E(iu),
\end{equation}
[recall Eq.~(\ref{greencavity1})], which vanishes at the cavity center.
For small cavity radii Eq.~(\ref{K}) reduces to
\begin{multline}
K=\frac{\hbar}{12\pi^2\varepsilon_0c^5} \int _0^\infty
\mathrm{d}u\,u^5\alpha_A(iu)
\left\{90\,\frac{\varepsilon(iu)-1}{3\varepsilon(iu)+2}\,
\frac{c^5}{u^5 R_\mathrm{c}^5}\right.\\
-\left.\frac{75}{7}\,\frac{\varepsilon^2(iu)[7\mu(iu)+3]
-6\varepsilon(iu)-4}
{[3\varepsilon(iu)+2]^2}\,\frac{c^3} {u^3 R_\mathrm{c}^3}\right\},
\end{multline}
leading to a force pointing away from the center of the cavity
(corresponding to an unstable equilibrium) for materials with
purely dielectric properties and to a harmonic force for purely
magnetic materials.


\end{document}